\documentclass[useAMS,usenatbib]{mn2e}

\usepackage{graphicx,amssymb,amsmath}

\title[Test of dynamical models with GCs data]{Testing multi-mass dynamical models of
star clusters with real data: mass segregation in three Galactic globular clusters\thanks{Based on FOcal Reducer and low dispersion
Spectrograph (FORS) observations collected with the Very
Large Telescope of the European Southern Observatory, Cerro Paranal, Chile,
within the observing programmes 091.D-0562(A) and 093.D-0228(A) (PI: Dalessandro), and on observations made with the
Large Binocular Camera at the Large Binocular Telescope within program
INAF.D07-2010A (PI: Ferraro) and 
with the Advanced Camera for Surveys (ACS) on board to the NASA/ESA Hubble Space Telescope under programs GO-10775 (PI: Sarajedini).}}
\author[Sollima et al.]{A. Sollima$^{1,2}$\thanks{E-mail:
antonio.sollima@oabo.inaf.it}, E. Dalessandro$^{1,2}$, G. Beccari$^{3}$, C. Pallanca$^{2}$\\
$^{1}$ INAF Osservatorio Astronomico di Bologna, via Ranzani 1, Bologna, 40127,
Italy\\
$^{2}$ Dipartimento di Astronomia, Universit{\'a} di Bologna, via Ranzani 1,
40127, Bologna, Italy\\
$^{3}$ European Southern Observatory, Karl-Schwarzschild-Strasse 2, D-85748
Garching bei Munchen, Germany\\
}
\begin{document}


\pagerange{\pageref{firstpage}--\pageref{lastpage}} \pubyear{2016}

\maketitle

\label{firstpage}

\begin{abstract}
We present the results of the analysis of deep photometric data for a sample of
three
Galactic globular clusters (NGC5466, NGC6218 and NGC 6981) with the 
aim of estimating their degree of mass segregation and testing the predictions of
analytic dynamical models.
The adopted dataset, composed by both Hubble Space Telescope and ground based 
data, reaches the low-mass end of the mass functions of these
clusters from the center up to their tidal radii allowing to derive their radial distribution of stars with different masses.
All the analysed clusters show evidence of mass segregation with the most
massive stars more concentrated than low-mass ones.
The structures of NGC5466 and NGC6981 are well reproduced by multimass 
dynamical models adopting a lowered-Maxwellian distribution function and the prescription for
mass segregation given by Gunn \& Griffin (1979). Instead, NGC6218 appears to be more mass
segregated than model predictions.
By applying the same technique to mock observations derived from 
snapshots selected from suitable N-body simulations we show that
the deviation from the behaviour predicted by these models depends on the 
particular stage of dynamical evolution regardless of initial conditions.
\end{abstract}


\section{Introduction}
\label{intro_sec}

The dynamical evolution of globular clusters (GCs) is one of the most intriguing
topics of stellar astrophysics. Indeed, GCs are stellar systems composed by
billions of stars subject to their mutual gravitational attraction being the best
representation in nature of the "gravitation N-body problem".
Moreover, since GCs are the oldest stellar systems known, 
their half-mass relaxation time is often shorter that their ages and processes 
like kinetic energy equipartition, mass segregation and core collapse can be 
at work and leave signatures in the phase-space distribution of their stars.

Stars in GCs cover a wide range of masses from $\sim0.1~ M_{\odot}$ (the
faintest Main Sequence stars) to $>14~M_{\odot}$ (the heaviest black holes;
Belczynski et al. 2010)
with a distribution which varies from cluster to cluster and is generally
bottom-heavy (with a larger number of low-mass stars than high-mass
ones; Paust et al. 2010). Because of the cumulative effect of many long-range two-body encounters, stars with
large kinetic energies (i.e. massive and/or fast) lose energy to less energetic
(low-mass and/or slow) ones. As a consequence, high-mass stars sink toward
less energetic orbits preferentially located in the innermost cluster region,
while low-mass stars diffuse in an extended halo. So, the variation
of the density and velocity dispersion distribution across the cluster extent
correlated with stellar mass represents the most direct evidence
of dynamically evolution of a star cluster.

Observational evidence of mass segregation have been shown for many GCs using as
a benchmark the radial variation of the mass function (MF; Da Costa 1982; Irwin
\& Trimble 1984; Pryor, Smith \& McClure 1986; Richer, Fahlman \& Vandenberg
1988; Bolte 1989; Fahlman, Richer \& Nemec 1991;
Drukier et al. 1993; Paresce, De Marchi \&
Jedrzejewski 1995; De Marchi \& Paresce 1996;
Ferraro et al. 1997; 
Zaggia, Piotto \& Capaccioli 1997; Fischer et al. 1998; 
Rosenberg et al. 1998; Saviane et al. 1998; Rood et al. 1999;  
Andreuzzi et al. 2000, 2004; 
Albrow, De Marchi \& Sahu 2002; Lee et al. 2003, 2004; 
Koch et al. 2004; Pasquali et al. 2004; 
Pasquato et al. 2009; Balbinot et al. 2009; 
Beccari et al. 2010; Goldsbury, Heyl \& Richer 2013; Martinazzi et al. 2014;
Frank, Grebel \& K{\"u}pper 2014; Zhang et al. 2015) or the radial distribution
of massive objects like Blue Straggler Stars and/or binaries (Dalessandro et al.
2015; Ferraro et al. 2012 and references therein).

On the other hand, kinematic evidence of energy exchange between mass groups 
represents an observational challenge because of the prohibitive performances
required to extract accurate kinematics for large samples of GCs stars in a
significant range of magnitudes along the Main Sequence (MS).
Indeed, radial velocities have been obtained essentially only for 
the brightest stars, giants and subgiants, which have very similar masses.
To date, the measure of a
significant number of radial velocities for stars in a range of masses
has been performed through integral field spectroscopy only in the GC NGC 6397,
leading however to uncertain results (Kamann et al. 2016).
The same kind of analysis made through proper motions requires superb
astrometric accuracies that have been achieved only recently using Hubble Space 
Telescope (HST) multi-epoch observations providing 
evidence for mass-dependent kinetic temperature in M15 (Bellini et al. 2014).

The presence and degree of mass segregation affect the determination of many
structural parameters like its mass, size, MF, mass-to-light ratio,
etc. These quantities are indeed estimated using information coming from the brightest (relatively 
massive) stars for which kinematics are available and contributing to the large 
majority of the cluster light, and later corrected using the predictions of
dynamical models. 

The most direct way to model the dynamical evolution of a stellar
system is through the use of N-body simulations.
However, in spite of the impressive progress of the computing power in the last decades,
the fit of star cluster observables with direct N-body simulations has been feasible only
for open cluster-like objects (Hurley et al. 2005; Harfst,
Portegies Zwart \& Stolte 2010) or small GCs (Zonoozi et al. 2011, 2014; Heggie 2014; Wang et
al. 2016). This is because a single 
simulation with a number of particles consistent with that observed in GCs
($10^{5-6}$) requires months of computing time, and a large number of simulations
are needed to tune the initial conditions in such a way to reproduce after a Hubble
time the present-day structure and properties of a given GC.
Similarly, Monte Carlo simulations, although providing a significant improvement in speed,
are subject to the same limitation (Giersz 2006; Giersz, Heggie \& Hurley 2008; 
Giersz \& Heggie 2009, 2011). 

An alternative way to model the structure of a multi-mass stellar system
is the use of analytic models.
Analytic models are generally defined by distribution functions depending on 
constants of the motion, and assume that the cluster is in a steady state and 
in equilibrium with the surrounding tidal field. Because star clusters are 
collisions-dominated systems, we have a relatively advanced 
understanding of the distribution of their stars in phase space from theory and 
numerical simulations, and the choice of distribution function-based models is 
justified. The most popular model of this kind is the King (1966) model which 
proved to be quite effective in reproducing the surface brightness profiles of 
many GCs, open clusters and dwarf galaxies (Djorgovski 1993; 
McLaughlin \& van der Marel 2005; Carballo-Bello et al. 2012; Miocchi et al. 2013). 
A generalization of this model accounting for radial anisotropy and a degree of 
equipartition among an arbitrary number of mass components has been provided by 
Gunn \& Griffin (1979; see also Da Costa \& Freeman 1976; Merritt 1981). 
The underlying assumptions of these models (i.e. the functional dependence of 
the distribution function on the integrals of motion and masses), although 
relying on a physical basis, are only arbitrary guesses to model the result of 
the complex interplay among many physical processes. It is therefore essential
to test the predictions of these models with suitable sets of observational data.

Since the first observational evidence of mass-segregation in GCs, the prediction of analytic 
multi-mass models have been compared with observations, providing generally good 
results (Richer \& Fahlman 1989; King, Sosin \& Cool 1995; Sosin 1997; 
Richer et al. 2004; Beccari et al. 2015).
However, these studies consist mainly of only two pointings at different
distances from the cluster center and sample a small portion of the cluster 
MF.
In the absence of suitable observational data, recent studies tested the
prediction of analytical models using snapshots extracted from sets of numerical 
simulations. In this regard, Takahashi \& Lee (2000) fitted the outcome of 
thier Fokker-Planck simulations including the effect of an external 
tidal field and anisotropy with multimass King-Michie models and found that
while they provide a reasonable fit of the more massive components, they
underestimate two-body relaxation effects for low-mass stars leading to
significant biases in the conversion between local and global MF.
Trenti \& van der Marel
(2013) noted that the simulated clusters never reach complete kinetic energy 
equipartition (confirming what already found by Inagaki \& Saslaw 1985 and 
Baumgardt \& Makino 2003 through Fokker-Planck and N-body simulations) and argued that the widely used King-Michie models could be 
inadequate to model these stellar systems. In Sollima et al. (2015) the biases
in the estimation of the 
mass and MF of two N-body simulations have been quantified by 
adopting different fitting techniques employing multi-mass King-Michie models. 
In that work, we found that these quantities are correctly estimated (within
$\sim$10\%) during the cluster evolution after the first half-mass
relaxation time.

\begin{figure*}
 \includegraphics[width=18cm]{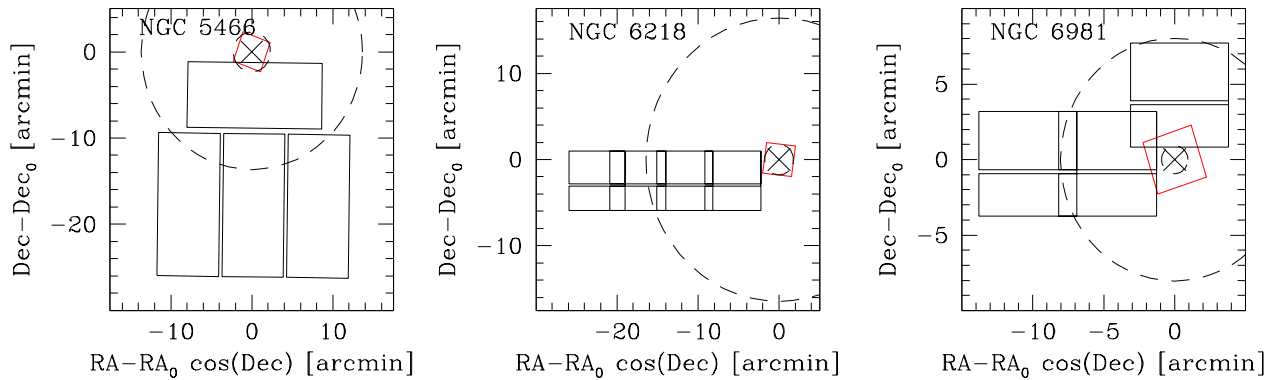}
 \caption{Maps of the region sampled by our observations. The ACS and ground based pointings are marked 
 with red (grey in the printed version of the paper) and black lines, respectively. The cluster center, the half-light and the tidal radii derived by
 McLaughlin \& van der Marel (2005) are shown as dashed lines.}
\label{map}
\end{figure*}

In this paper we present an in-depth analysis of the radial distribution of
stars in a relatively wide range of masses in three Galactic GCs, namely NGC
5466, NGC 6218 and NGC 6981. The aim of this work is to quantify the accuracy
of multi-mass models in reproducing the degree of mass segregation measured in these GCs.
In Sect. 2 we describe the adopted photometric dataset together with the
observational strategy and reduction technique. Sect. 3 is devoted to the
description of the fitting algorithm. The results of the application of the 
adopted technique to a set of N-body simulations and the comparison with
observations is shown in Sect. 4. Finally, we summarize our results in Sect. 5.

\section{Observational material}
\label{obs_sec}

\begin{table}
 \centering
 \begin{minipage}{140mm}
  \caption{Observing logs.}
  \begin{tabular}{@{}lcccr@{}}
  \hline
  Name & Instrument & filter & Exp. time  & \# of exp.s\\
       &            &        &  (sec)         & per field   \\
 \hline
NGC 5466 & LBC   & B & 5  & 1\\
  &	  & B & 90 & 7\\
  &	  & B & 400 & 11\\
  &	  & V & 5  & 1\\
  &	  & V & 60 & 7\\
  &	  & V & 200 & 15\\
 \hline
NGC 6218 & FORS2 & V & 165 & 15\\
  &   & I & 105 & 12\\
 \hline
NGC 6981 & FORS2 & V & 515 & 8\\
  &   & I & 240 & 16\\
\hline

\end{tabular}
\end{minipage}
\end{table}

\begin{figure*}
 \includegraphics[width=18cm]{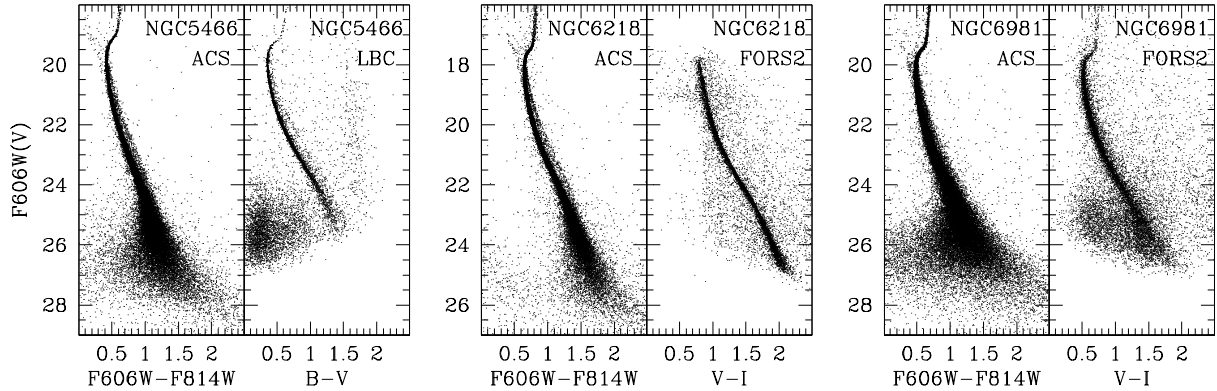}
 \caption{CMDs of the three analysed GCs. In each panel both the ACS (left inset)
 and ground-based (right inset) are shown.}
\label{cmd}
\end{figure*}

The main photometric database used to sample the central part of all the target 
GCs is constituted by the set of publicly 
available deep photometric catalogs of the 
"globular cluster treasury project" (Sarajedini et al. 2007). It consists of 
high-resolution HST images secured with the Advanced 
Camera for Surveys (ACS) Wide Field Channel through the F606W and F814W filters. 
The field of view of the camera ($202\arcsec \times 202\arcsec$) is centered 
on the clusters' center with a dithering pattern to cover the gap between the 
two chips, allowing a full coverage of the core of the GCs considered in 
our analysis. This survey provides deep color-magnitude diagrams (CMDs) 
reaching the faint MS of the target clusters down to the 
hydrogen burning limit (at $M_{V}\sim 10.7$) with a signal-to-noise ratio 
S/N$>$10. The results of artificial star experiments are also available to 
allow an accurate estimate of the completeness level and photometric errors. 
A detailed description of the photometric reduction, astrometry, and artificial 
star experiments can be found in Anderson et al. (2008). 

The outer region of NGC 5466 has been analysed using images collected
with the Large Binocular Camera (LBC) mounted at the Large Binocular Telescope 
(Mount Graham, Arizona) while for the other two GCs 
the FOcal Reducer and low dispersion Spectrograph 2 (FORS2) camera at the Very Large Telescope 
of the European Southern Observatory (ESO; Cerro Paranal, Chile) has been used. 
Maps of the ACS, LBC and FORS2 pointings across the GCs field of view are shown
in Fig. \ref{map}.

The LBC is a wide-field imager which provides an effective
$23\arcmin\times23\arcmin$ field of view, sampled at 0.224
arcsec/pixel over four chips of 2048$\times$4608 pixels each.
Observations were preformed in a photometric night (April 11th 2010) using the blue detector of the LBC
camera. A set of short exposures was secured in the B and V filters with the cluster 
center positioned in the central chip of the LBC-blue CCD mosaic. Deep images 
were obtained with the LBC-blue FOV positioned $\sim100\arcsec$ south from the 
cluster center. This dataset has been recently used to derive the
radial distribution of Blue Stragglers and binaries as well as the MF
of NGC 5466 (Beccari et al. 2013, 2015).

The FORS2 camera has been used with the Standard Resolution collimator
providing a pixel scale of 0.25 arcsec/pixel and a field sizes of
$6.8\arcmin\times6.8\arcmin$. A set of deep V and I images has been secured in a mosaic
pattern allowing to reach the nominal tidal radii of the observed clusters. 
Raw images were corrected for bias and flat field, and the overscan region was 
trimmed using the standard {\rm IRAF} tasks. 

The average seeing was comprised in the interval $0.\arcsec 5-1.\arcsec 0$
remaining stable within $0.\arcsec 2$ during each observing night so that all
the images of both datasets were used in the photometric analysis.
A log of the observations is listed in Table 1.

The raw LBC and FORS2 frames were corrected for bias and flat fields and the
overscan region was trimmed using LBC Survey data center pipeline and the 
{\rm IRAF}\footnote{IRAF is distributed by the National Optical Astronomy Observatories,
which are operated by the Association of Universities for Research
in Astronomy, Inc., under cooperative agreement with the National
Science Foundation.} task {\it imred}, respectively.
The photometric reduction of both datasets have been performed using the
DAOPHOT/ALLFRAME PSF-fitting routine (Stetson 1994).
Images were aligned and corrected for geometric distorsion using a
third-order polynomial. We performed the source detection on the stack of all images while the 
photometric analysis was performed independently on each undistorted image. Only stars 
detected in two out of three long exposures or in the short ones have been 
included in the final catalog. We used the most isolated and brightest stars in 
the field to construct a PSF which has been modelled as a Moffat function
plus a numerical component which varies quadratically across the field of view. 
The same PSF stars were also employed to link the aperture magnitudes to the instrumental 
ones (a single magnitude shift has been calculated for each chip). 
Instrumental magnitudes have been then transformed into the standard
Johnson-Cousin photometric system 
using a first order (zero point + color term) linear relation obtained by comparing the stars in common
with the standard fields by Stetson\footnote{http://www.cadc-ccda.hia-iha.nrc-cnrc.gc.ca/en/community/STETSON/standards/}. The final catalogs have been astrometrically calibrated 
through a cross-correlation with the Two Micron All Sky Survey catalog (Skrutskie et al. 2006). 
The astrometric solution has a typical standard deviation of 200 mas.
A detailed description of the photometric reduction procedure can be
found in Beccari et al. (2013) for the LBC data and in a forthcoming paper (Dalessandro et al.,
in prep.) for the FORS2 ones.

The final CMDs sample the entire unevolved population of the analysed clusters 
down to $\sim$7 mag below the turn-off (see Fig. \ref{cmd}). Artificial stars
experiments have been performed using the procedure described in Bellazzini et
al. (2002). A list of input positions and magnitudes has
been produced by distributing artificial stars in random positions within a grid
of cells (one star per cell) homogeneously distributed across the
field of view. The V magnitude of artificial stars has been extracted from a power-law
distribution with index x=-2 while their colors have been obtained by 
interpolating on the cluster ridge line. Artificial stars have been then added
to the science frames using the corresponding PSF.
The photometric analysis has been then repeated using the same procedure adopted
for the science frames producing a catalog of output positions and magnitudes
for artificial stars. At the end of the above procedure a catalog of 100,000
artificial stars have been produced allowing a proper estimate of the
photometric accuracy and the completeness level at different magnitudes 
across the observed field of view.

\section{Method}
\label{met_sec}

We tested the agreement between King-Michie analytic models and observations by
comparing the distribution of star masses as a function of projected distances from the cluster 
center. In the following sections we describe the technique adopted to derive masses from our
photometric dataset as well as the adopted models and the fitting algorithm.

\subsection{Stellar Masses Estimate}
\label{stell_sec}

GC stars occupy different regions of the CMD
according to their evolutionary stages and masses, so an estimate of their masses can be made by means
of the comparison with suitable theoretical isochrones. 
We adopted the theoretical isochrones by Dotter et al. (2007) with metallicity
[Fe/H]=-2.31, -1.33 and -1.48 (for NGC5466, NGC6218 and NGC6981,
respectively; Carretta et al. 2009) and appropriate ages chosen by best-fitting the morphology of
the turn-off region of the observed CMD (13.2, 13 and 12.3 Gyr, respectively, in
agreement with those estimated by Dotter et al. 2010). Absolute magnitudes have been
converted into the observational color-magnitude plane assuming the distance moduli
and reddening listed in the Harris catalog (Harris 1996; 2010 edition).

\begin{figure}
 \includegraphics[width=8.6cm]{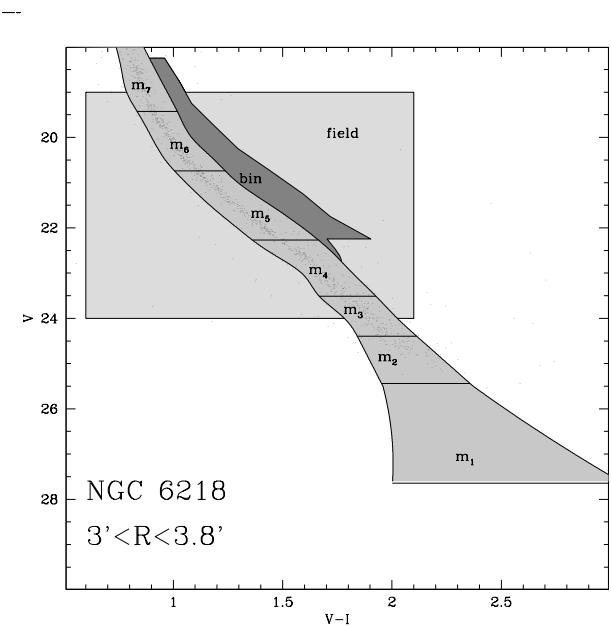}
 \caption{Selection boxes adopted for the population of single stars ($m_{1}$ to
 $m_{7}$), binaries (bin) and field stars (field) of NGC 6218. The (V, V-I) CMD 
 is overplotted.}
\label{box}
\end{figure}

This task is however complicated by the contamination from fore/background 
field stars and by the effect of unresolved binaries and photometric
errors spreading out stars far from their original location in the CMD,
making difficult to assign them a proper mass. 
Although it is not possible to unambiguously distinguish binaries, single and
field stars across the entire CMD, we adopt a statistical classification of cluster
members. 
In particular, we adopted the following procedure:
\begin{itemize}
\item{The field of view of the photometric data has been divided in
annular concentric regions. We defined 16 regions with both width and
separation of $0.\arcmin 1$ in the
ACS field of view and 8 regions separated by logarithmic steps of 0.1 starting from 3$\arcmin$ for the
ground based data.} 
\item{For
each region a synthetic CMD (containing $\sim10^{6}$ stars) has been simulated by randomly extracting masses from a 
power-law MF and derived the corresponding magnitudes by
interpolating through the adopted isochrone.
A population of binaries has been
also simulated by associating to a fraction of stars a secondary
component whose mass has been randomly extracted assuming a flat mass-ratios 
distribution (Milone et al. 2012). The fluxes of the two components have been
then summed in both passbands to derive their corresponding magnitudes and color.
A synthetic population of field stars has been also added using the Galactic model by Robin et al. (2003)
whose magnitudes have been converted into the ACS photometric system using the transformations by Sirianni
et al. (2005).

In each annulus, the MF slope, the binary fraction and the number of field stars have been tuned to reproduce
the observed relative ratios of number counts in nine regions of the CMD defined as
follows (see Fig \ref{box}): 
\begin{itemize}
\item{seven V (F606W) magnitude intervals were defined corresponding to 
equal-mass intervals and including all stars with colors 
within three times the photometric error corresponding to their magnitudes;} 
\item{a region including the bulk of the binary populations with high mass ratios
($q>0.5$). This last region is delimited in magnitudes by the loci of binaries with primary 
star mass $M_{1} = 0.45 M_{\odot}$ (faint boundary) and $M_{1} = 0.75
M_{\odot}$ (bright boundary), and in color by the MS ridge line (blue boundary) 
and the equal-mass binary sequence (red boundary), both redshifted by three 
times the photometric error;}
\item{a region including mainly field stars in a magnitude interval between 1
and 5 mag in the V band below the turn-off and within the color range $0.6<V-I<2.1$.}
\end{itemize}
}
\item{For each synthetic star a particle in the same radial range and 
with magnitudes within 0.25 mag has been extracted  
from the library of artificial stars and, if
recovered\footnote{An artificial star has been
considered recovered if its input and output magnitudes differ by less than 2.5
log(2) ($\sim0.75$) mag in both F606W and F814W magnitudes.}, the magnitude and 
color shift with respect to its input quantities 
has been added to those of the corresponding star;}
\item{As a final step, we
associated to each observed star the mass and the classification flag (single,
binary or field contaminant) of the closest particle in 
the synthetic CMD. }
\end{itemize} 

We limited our analysis to stars fainter than the turn-off since in the
ground-based dataset most of the bright stars belonging to the evolved 
population saturate in the deep exposures. These objects constitute less than
5\% of the total cluster population, so their exclusion does not affect the 
results of the present analysis. 

\subsection{Analytic Models}
\label{mod_sec}

We fitted our dataset with a set of King-Michie analytic models (Gunn \& Griffin 1979).
These models are constructed from a lowered-Maxwellian distribution function
made by the contributions of H mass groups
\begin{eqnarray*}
f(E,L)&=&\sum_{j=1}^{H}k_{j} f_{j}(E,L,m_{j})\nonumber\\   
f_{j}(E,L,m_{j})&=&exp\left(-\frac{A_{j}L^{2}}{2\sigma_{K}^{2}r_{a}^{2}}\right)
\left[ exp\left(-\frac{A_{j}E}{\sigma_{K}^{2}}\right)-1 \right]\nonumber\\
\end{eqnarray*}
where $E$ and $L$ are, respectively, the energy and angular momentum per unit mass, 
 $r_{a}$ is the radius beyond which orbits become
biased toward the radial direction, $A_{j}$ and $k_{j}$ are scale factors for each mass group 
and $\sigma_{K}$ is a normalization term. 
Although the distribution
function defined above allows for various levels of radial anisotropy 
we used only isotropic 
models in our comparison (assuming $r_{a}=+\infty$). This is because {\it i)} the presence of a significant degree
of velocity anisotropy cannot be verified without a large set of kinematic data
(see e.g. Watkins et al. 2015), and {\it ii)} the mass-dependence of anisotropy
in these models is oversimplified and it does not properly reproduce the 
behaviour evidenced in N-body simulations in many stages of their evolution 
(Sollima et al. 2015).
The dependence on mass of the 
coefficients $A_{j}$ determines the degree of mass segregation of the cluster.
In the formulation by Gunn \& Griffin (1979) $A_{j}\propto m_{j}$ which implies
that more massive stars are kinematically colder than less massive ones.
Under these assumptions, the distribution function of each mass group is
uniquely a function of energy and can be written as a function of the distance
from the center, the velocity and the star mass.
\begin{equation}
\label{eq_df}
f_{j}(v,r,m_{j})=exp\left(-\frac{m_{j}v^{2}}{2
m_{K}\sigma_{K}^{2}}-\frac{m_{j}\psi(r)}{m_{K}\sigma_{K}^{2}}\right)-1
\end{equation}
where $\psi$ is the effective potential defined as the difference between the cluster 
potential $\phi$ at a given radius $r$ and the potential at the cluster tidal radius 
($\psi \equiv \phi - \phi_{t}$) and $m_{K}$ is an arbitrary constant with the dimension of a mass.

The 3D number density of
each mass group are obtained by integrating Eq. \ref{eq_df} over the velocity domain
\begin{eqnarray}
\label{eq_den}
\rho_{j}(r)&=&\int_{0}^{\sqrt{-2\psi(r)}}4\pi v^{2} k_{j} f_{j}(v,r,m_{j}) dv\nonumber\\
\rho(r)&=&\sum_{j=1}^{H} \rho_{j}(r)
\end{eqnarray}
while the potential 
at each radius is determined by the Poisson equation
\begin{equation}
\label{eq_poiss}
\nabla^{2}\psi=4\pi G \rho
\end{equation}

Equations \ref{eq_den} and \ref{eq_poiss} have been integrated after assuming
as a boundary condition a value of the potential and its derivative at the
center ($\psi_{0};~d\psi/dr(0)=0$) outward till the
radius $r_{t}$ at which both density and potential vanish (see Gunn \& Griffin
1979 and Sollima et al. 2015 for a detailed derivation of the model properties).

Note that the MF is univocally linked to the $k_{j}$ coefficients, since the
number of stars in a given mass bin is given by
$$N_{j}=16 \pi^{2} k_{j} \int_{0}^{r_{t}} r^{2} \int_{0}^{\sqrt{-2\psi(r)}} v^{2}
f_{j}(v,r,m_{j})~dv~dr$$
So, the shape of the density profiles 
are completely determined by the parameters ($\psi_{0},~N_{j}$) while the
constant $m_{K}\sigma_{K}^{2}$ determines the mass of the model and the core
radius $r_{c}\equiv\sqrt{\frac{9\sigma_{K}^{2}}{4\pi G \rho_{0}}}$ gives the
size of the system.

As a last step, the above profiles have been projected onto the plane of the sky 
to obtain the surface density of each bin.

$$\Gamma_{j}(m_{j},R)=2\int_{R}^{r_{t}}\frac{\rho_{j} r}{\sqrt{r^{2}-R^{2}}}~dr$$

Here we considered 7 evenly spaced mass 
bins ranging from 0.1 $M_{\odot}$ to the mass at the RGB tip ($M_{tip}$), plus an additional
bin containing all the stellar objects more massive than $M_{tip}$ (like heavy binaries and
white dwarfs).

\subsection{Fitting algorithm}
\label{fit_sec}

The best-fit model has been chosen as the one providing the largest value of the 
merit function defined as follows
\begin{equation}
L=\sum_{i=1}^{N_{s}} log[P(m_{i},R_{i})]
\label{eq_like}
\end{equation}

where $N_{s}$ is the total number of stars flagged as singles and $P(m_{i},R_{i})$ is the probability 
density to find a single star with mass $m_{i}$ at a
projected distance $R_{i}$ from the cluster center according to a given model.
This last quantity can be calculated as
$$P(m_{i},R_{i})=\frac{R_{i}~X_{s}(m_{i},R_{i})}{\sum_{j=1}^{H} \mu_{s}(m_{j})~\Delta m~\int_{0}^{R_{max}}R~X(m_{j},R)~dR}$$
and
\begin{equation}
\label{x_eq}
X_{s}(m,R)=\mu_{s}(m)~\Gamma(m,R)~C_{s}(m,R)~A(R)
\end{equation}

where $C_{s}(m,R)$ is the mean completeness estimated for stars with mass
$m$ at a projected distance $R$ from the cluster center, $A(R)$ is the
azimuthal coverage of the observational field of view at a given projected
distance, $\mu_{s}(m)$ is the fraction of single stars with a given mass, $\Delta m$
is the mass spacing between the model bins and $R_{max}$ is the maximum distance
where the fit is performed. 

The completeness factor as a function of mass and distance $C(m,R)$ has been calculated 
for both singles and binaries by estimating the V and I magnitude corresponding 
to the given mass. We adopted the mass-luminosity relation of the best-fit
isochrone for single stars while for binaries the average magnitudes are
calculated from
the synthetic CMD simulated as described in Sect. \ref{stell_sec}. The completeness
has been then 
calculated as the fraction of recovered stars in the artificial star
catalog with input magnitudes 
within 0.25 mag in both bands and in a distance range within $0.\arcmin 05$ from
the given projected radius.

The function $\mu_{s}(m)$ allows to account for the population of binaries and of 
dark remnants
contributing to the mass budget of the cluster but not contained in our sample.
This parameter is defined as 
\begin{equation}
\mu_{s}(m)\equiv 1-\mu_{remn}(m)-\mu_{b}(m)
\label{eq_mu}
\end{equation} 
where $\mu_{b}(m)$ and $\mu_{remn}(m)$ are the
fraction of binaries and remnants at a given mass, respectively.
These quantities have been calculated in each step of the fitting algorithm (see
below) starting from assumptions on the global binary fraction ($f_{b}$) and the
MF slope ($\alpha$). In particular:
\begin{itemize}
\item{We simulated a synthetic population of stars extracted from a Kroupa
(2001) MF between 0.1 $M_{\odot}$ and 8 $M_{\odot}$;}
\item{We simulated a population of binaries by associating to a fraction $f_{b}$ 
of these stars a companion star extracted from a flat distribution of mass ratios;}
\item{We evolved passively all stars (singles and binaries) using the
prescriptions of Kruijssen (2009). Because of the upper limit of the adopted initial MF,
no neutron stars and black holes are present, consistently with the
expectation that they are ejected at birth in these low-mass stellar systems
(Kruijssen 2009);}
\item{A fraction of stars are removed as a function of their mass to simulate
the effect of evaporation. We assumed that stars evaporate with 
efficiencies which are a function of their masses only, and are set by the ratio of
the present-day MF and the Kroupa (2001) initial MF.
For this purpose a random number $\epsilon$ has been
extracted from a uniform distribution between 0 and 1 and the star is retained
if $\epsilon<Ret(M)$ where
$$
Ret(M)=\begin{cases} 
\left(\frac{M_{tip}}{0.5
M_{\odot}}\right)^{\alpha-2.3}~\left(\frac{M}{0.5
M_{\odot}}\right)^{1.3-\alpha}& \mbox{if } M\leq 0.5 M_{\odot}\\
\left(\frac{M}{M_{tip}}\right)^{2.3-\alpha}& \mbox{if } 0.5
M_{\odot}<M\leq M_{tip}\\
1 & \mbox{if } M>M_{tip}
\end{cases}$$
}
\item{The values of $\mu_{s}$, $\mu_{bin}$ and $\mu_{remn}$ have been calculated as the
fraction of singles, binaries and remnants in the final population in different mass bins}
\end{itemize}

We set the value of $R_{max}$ to the distance where a power-law
density profile starts to develop. This feature is in fact a consequence of
deviation from equilibrium occurring because of the interaction with the Milky
Way tidal field (Johnston, Sigurdsson \& Hernquist 1999; Testa et al. 2000; 
K{\"u}pper et al. 2010) and cannot be properly
accounted by King-Michie models assuming equilibrium {\it a priori}.

To search the best model in reproducing the distribution of stars in the (m, R)
plane an iterative algorithm has been emplojed.
We start from an initial guess of the MF (through the coefficients $N_{j}$) and
of the global binary fraction $f_{b}$ and
repeated the following step until convergence:
\begin{itemize}
\item{The slope of the MF $\alpha$ is derived by fitting the $N_{j}$
coefficients with a power-law. The corresponding $\mu_{s}(m)$ and $\mu_{b}(m)$
functions are calculated using eq. \ref{eq_mu};}
\item{The optimal value of $\psi_{0}$ and $r_{c}$ are searched using a 
Markov Chain Monte Carlo
algorithm maximizing the merit function defined in eq. \ref{eq_like};}
\item{The relative fraction of stars in the eight mass bins $N_{i}$ and the
global binary fraction $f_{b}$ are adjusted 
by multiplying them for corrective terms which are proportional to the ratio 
between the relative number counts in each bin of the observed sample and the
corresponding model prediction
\begin{eqnarray*}
N_{j}'&=&N_{j}\left(\frac{N_{j}^{obs}}{N_{s}~\Delta
m\int_{0}^{R_{max}}P(m_{j},R)~dR}\right)^{\eta}\nonumber\\
f_{b}'&=&f_{b}\left(\frac{N_{bin}^{obs}}{N_{s}}
\frac{\sum_{j=1}^{H}\int_{0}^{R_{max}}X_{s}(m_{j},R)~dR}
{\sum_{j=1}^{H}\int_{0}^{R_{max}}X_{b}(m_{j},R)~dR}
\right)^{\eta}\nonumber
\end{eqnarray*}
where $N_{j}^{obs}$ is the number of stars observed in the j-th mass bin, 
$N_{bin}^{obs}$ is the number of stars flagged as binaries, $X_{b}$ is the same
function defined in eq. \ref{x_eq} but using the functions $\mu_{b}$ and $C_{b}$
for the binary population and
$\eta$ is a softening parameter, set to 0.5, used to avoid divergence.}
\end{itemize}
The above procedure converges after $\sim$10 iterations providing the
combination of parameter ($\psi_{0},~r_{c},~N_{j},~f_{b}$) corresponding to the maximum
likelihood.

\section{Results}
\label{res_sec}
 
\begin{figure}
 \includegraphics[width=8.6cm]{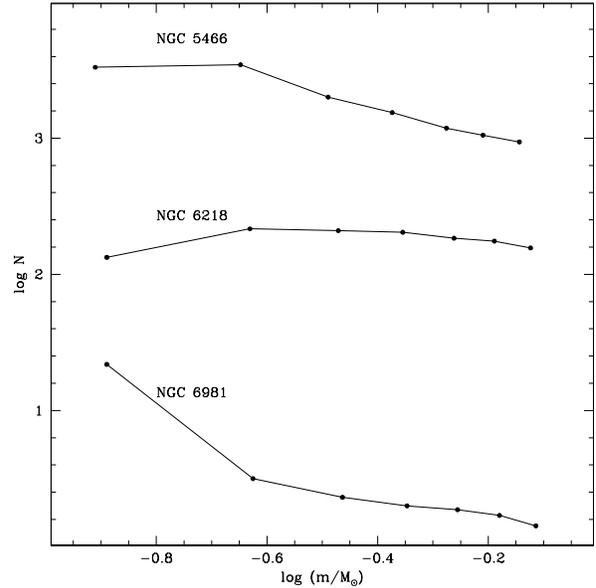}
 \caption{Modelled present-day global MFs of the three target clusters. 
 A vertical shift has been added to each cluster for clarity.}
\label{mf}
\end{figure}

\begin{table}
 \centering
 \begin{minipage}{140mm}
  \caption{Parameters of the best-fit models.}
  \begin{tabular}{@{}lccccr@{}}
  \hline
  Name & $log (M/M_{\odot}$) & $r_{h}$ & $\alpha$ & $f_{b}$ & $\beta$\\
       &                     & pc      &          &   \%    &\\
 \hline
NGC 5466 & 4.65  & 13.94 & -0.97 & 8.0 &  0.00$\pm$0.03 \\
NGC 6218 & 4.86  &  4.74 & -0.36 & 5.4 & -0.24$\pm$0.04\\
NGC 6981 & 4.81  &  7.70 & -0.55 & 5.8 &  0.07$\pm$0.04\\
\hline
NGC 6981remn & 4.88  &  7.89 & -0.55 & 6.4 &  0.09$\pm$0.03\\
NGC 6981bin  & 4.78  &  7.59 & -0.67 & 15.1 &  0.13$\pm$0.02\\
\hline
\end{tabular}
\end{minipage}
\end{table}

\begin{figure*}
 \includegraphics[width=18cm]{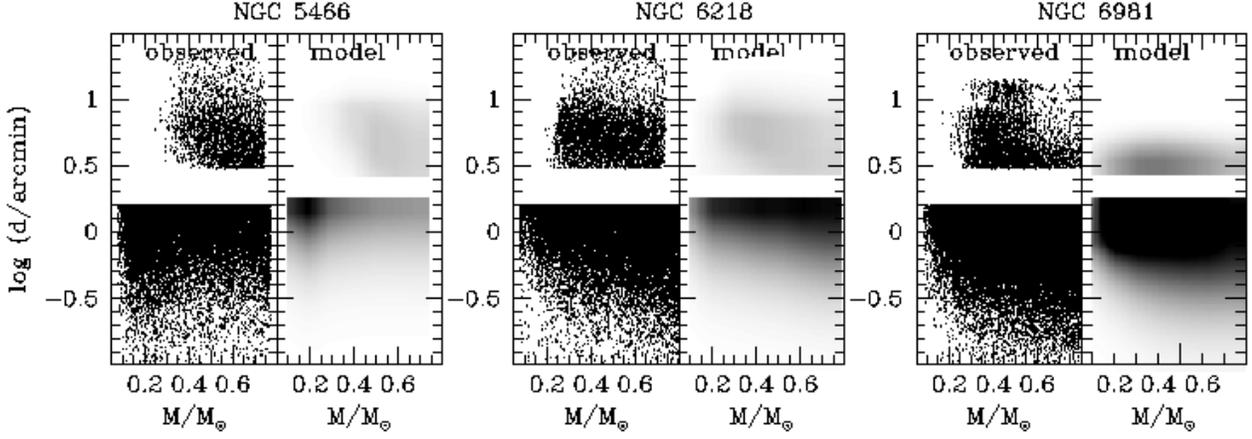}
 \caption{Distribution of single stars in the m-log R plane for the three target
 clusters (left panels). The probability density predicted by the best-fit
 models are shown in right panels.}
\label{res}
\end{figure*}

\begin{figure*}
 \includegraphics[width=18cm]{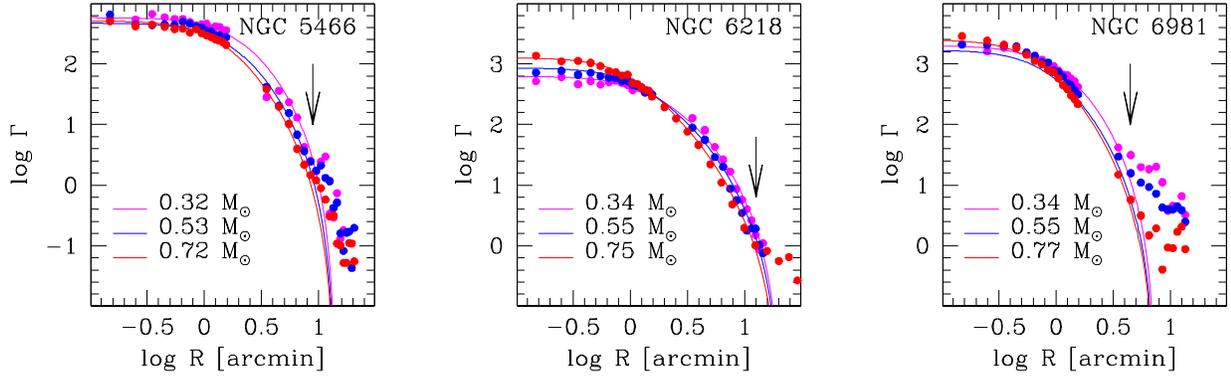}
 \caption{Completeness corrected density profiles for different mass groups in
 the three target clusters. The prediction of the best-fit King-Michie models are
 overplotted as solid lines. The outer limiting radii adopted in the fitting routine
 ($R_{max}$) are marked by vertical arrows.}
\label{prof}
\end{figure*}

The parameters of the best-fit King-Michie models are listed in Table 2 for the
three analysed clusters.

The estimated masses agree within the errors, although they are slightly smaller
by $\Delta logM\sim0.1-0.2$, with those listed 
by McLaughlin \& van der Marel (2005). Note that these authors calculated 
masses by fitting the
surface brightness profiles of a large sample of GCs with single-mass King 
(1966) models. So, some differences are expected because of the inadequacy of
single-mass models to reproduce the structure of clusters with steep MF 
because of
the significant contribution of faint low-mass stars to the total mass budget
without significantly affecting the surface brightness profile (see Sollima et
al. 2015). 

In Fig. \ref{mf} the derived global MFs of single stars are shown. 
Their corresponding slopes, listed in Table 2, have been calculated by fitting a power-law through
the 5 mass bins with $m_{j}>0.3 M_{\odot}$. The less massive bins are indeed
those prone
to the largest systematic uncertainties (see Sect. \ref{syst_sec}) and often show a profile deviating from a 
power-law behaviour. The derived values of the slopes for NGC5466 and NGC6218
agree with the estimates made by Paust et al. (2010), Beccari et al. (2015) and Sollima, Bellazzini \& Lee
(2012) while
for NGC6981 no previous estimate of the MF slope is present in the
literature.

A cluster-to-cluster comparison between the fraction of binaries estimated here and
those listed in Milone et al. (2012) indicate no significant differences, although
they are all systematically smaller than those estimated by these authors by 
$\Delta f_{b}\sim1.8-5.2\%$. Consider that the analysis 
of Milone et al. (2012) is conducted on the ACS field of view only, sampling the
central portion of the GCs where binaries tend to sink because of mass
segregation. This can explain the difference observed between these two works.

The comparison between the distribution of single stars in the $m-log~R$ plane for
the three analysed clusters and the prediction of the corresponding best-fit
models is shown in Fig. \ref{res}. Qualitatively, the density contours of the models 
well reproduce the distribution of stars in this plane for all clusters.
Another way to visualize such a comparison is shown in Fig. \ref{prof} where the
completeness corrected density profiles of stars in three different mass bins
are compared with the prediction of the best-fit models. Also in this case, the
agreement is generally good. It is worth noting that the models tend to
underpredict the density of all mass groups in the outermost portion of all
clusters at distances $R>R_{max}$. In this region, strong deviations from
equilibrium are apparent as power-law tails extending up to the GCs' tidal radii
and beyond. This effect is particularly evident in NGC6981 where such features
cover almost the entire extent of FORS2 fields.

To quantify the adequacy of models in reproducing the effect of mass segregation
we compared the cumulative radial profile of observed stars in the different 
mass bins ($F_{j}^{obs}(<R)$) with those predicted by models calculated through the relation
$$F_{j}^{mod}(<R)=\frac{\int_{0}^{R} R~P(m_{j},R)~dR}{\int_{0}^{R_{max}} R~P(m_{j},R)~dR}$$

For each bin we calculated the
logarithmic shift in radius $\Delta log r_{c,j}$ to be added to the model profile
to minimize the Kolmogorov-Smirnov statistic 
$$KS(m_{j})=max|F_{j}^{obs}(<R)-F_{j}^{mod}(<R)|$$
The slope of the $\beta=d \Delta log r_{c,j}/d log~m_{j}$ is then calculated and
adopted as an indicator of the adequacy of the best-fit King-Michie model in
reproducing the actual degree of mass segregation. Values of $\beta$ close to
zero indicate a good agreement between models and observation while positive
(negative) values indicate an overestimate (underestimate) of mass segregation.
Note that this approach is
similar to that adopted by Goldsbury et al. (2013) who quantify mass segregation
in a sample of GCs using the power-law slope of the half-power radius vs. mass
relation. In this case, the index $\beta$ defined above is almost
equivalent to the difference between the observed Goldsbury et al. (2013) slopes
and those of the corresponding best-fit models. The values of the index $\beta$
calculated for the three analysed clusters are listed in the last columns of
Table 2. While the values of $\beta$ for NGC5466 and
NGC6981 are small ($|\beta|<0.1$), a negative value of $\beta=-0.24\pm0.04$ is
measured for NGC6218 indicating that it is more mass segregated than what
predicted by the adopted model.

\subsection{Effect of assumptions}
\label{syst_sec}

In the algorithm described in Sect.s \ref{fit_sec} and \ref{stell_sec} we
adopted many assumptions. Among them, the prescription for the retention
fraction of remnants and the choice of the distribution of mass ratios in
binaries are arbitrary and not well constrained by observations. It is therefore
important to check the actual impact of such assumptions on the results
presented in Sect. \ref{res_sec}. For this purpose, as a test case, we repeated 
the analysis for NGC6981 by
assuming extreme conditions for these parameters: {\it i)} a 100\% retention of
all remnants (including neutron stars and black holes with progenitor masses up
to 120 $M_{\odot}$; named NGC6981remn in Table 2), and {\it ii)} a mass ratio
distribution of binaries drawn by randomly associating stars extracted from the
MF (corresponding to a distribution peaked at $q\sim0.3$ with a decreasing tail at large mass
ratios; NGC6981bin). The resulting masses, half-mass radii, MF slopes and binary
fractions are also listed in Table 2. It can be noted that masses
are significantly affected only by the choice of the retention fraction of remnants, while the MF 
slopes and binary fractions are sensitive only to the mass ratios distribution of binaries.
On the other hand, the values of $\beta$, quantifying the agreement between data
and models, are shifted toward positive values.
The effect of remnants on mass and size is due to the large masses of
these objects with respect to single stars. So, an adopted larger fraction of 
remnants provides an additional mass budget which is modelled as a
concentrated distribution of mass.

The large impact of the binary mass ratios distribution in the derived
binary fraction and present-day MF is due to the fact that 
this parameter affects the classification of single and binaries (see sect.
\ref{stell_sec}). In particular,
in case NGC6981bin a smaller fraction of binaries is located in the binary
selection box with respect to the standard assumption of a flat mass ratios
distribution. Thus a large fraction of binaries is needed to explain the
observed number of objects in this region of the CMD. 
Moreover, the mass ratios distribution affects also the distribution in magnitude
of objects classified as binaries. 
As a consequence, the MF of the remaining stars classified as singles turns out
to be also affected by this choice, in particular in the low-mass range. Note
that the magnitude of the distorsion of the MF introduced by the
particular choice of 
the mass ratio distribution of binaries exceeds by more than an order of magnitude the statistical
fluctuations due to Poisson noise.

It is interesting to note that the parameter $\beta$ increases with increasing
the mass contained in remnants or binaries. This means that the larger is the
fraction of mass contained in massive objects the larger is the degree of mass
segregation predicted by multimass models, becoming larger than what
observed. This occurs because a large population of massive
stars increases the mass contrast with respect to the average stellar mass
mimicking the effect of a steeper MF (see Sect. \ref{nbody_sec}).

\subsection{Comparison with N-body simulations}
\label{nbody_sec}

The result presented in the previous section indicates that, for the
adopted recipies on the dark remnant retention fraction and the mass ratios
distribution of binaries, King-Michie models well
reproduce the distribution of masses in two out of three GCs analysed here while
underestimating the actual degree of mass segregation in NGC6218.
It is worth noting that this last cluster has the shortest half-mass relaxation time
among the GCs of our sample and is therefore expected to be dynamically more evolved
than the other analysed clusters.
Indeed, as introduced in Sect. \ref{intro_sec}, mass segregation is expected to develop
and grow as a result of the increasing efficency of two-body relaxation. 
So, clusters in different stages of dynamical evolution are expected to be 
characterized by different degrees of mass segregation.
It is interesting to check whether the recipy for mass segregation of 
multimass King-Michie models is adequate in different stages of dynamical
evolution.
For this purpose we analysed different snapshots of two
N-body simulations of star clusters in different stages of their evolution 
and under different initial conditions.

The N-body simulations considered here have been performed using the 
collisional N-body codes {\rm NBODY4} and {\rm NBODY6} (Aarseth 1999) and are 
part of the surveys presented by Baumgardt \& Makino (2003) and Lamers,
Baumgardt \& Gieles 
(2013). Each simulation contains 131 072 particles with no primordial binaries. 
Particles were initially distributed following a King (1966) model with central 
dimensionless potential W0 = 5, regardless of their masses. 
The two simulations 
start with different half-mass radii (with $r_{h}$ = 1 and 11.5 pc, hereafter 
referred to as W5rh1R8.5 and W5rh11.5R8.5, respectively). Particle masses are
extracted from a Kroupa (2001) MF with a lower mass limits of 0.1 $M_{\odot}$ and an 
upper mass limit of 15 and 100 $M_{\odot}$, for the W5rh11.5R8.5 and 
W5rh1R8.5 simulation, respectively. In these configurations, 
the total cluster masses are 71236.4 $M_{\odot}$ and 83439 $M_{\odot}$ for
simulations W5rh1R8.5 and W5rh11.5R8.5, respectively. 
The cluster moves within a logarithmic 
potential having circular velocity $v_{circ} = 220 ~km~s^{-1}$, on a circular orbit 
at a distance of 8.5 kpc from the galactic centre. The corresponding initial 
Jacobi radius is $r_{J}$ = 61.15 pc, i.e. equal to the initial tidal radius of the W5rh11.5R8.5 
simulation. Because of their different Roche lobe filling factors, the tidal 
field affects the two simulations in extremely different ways. Moreover, the 
initial half-mass relaxation time is significantly longer in model W5rh11.5R8.5 
($t_{rh}$ = 4.97 Gyr) with respect to model W5rh1R8.5 ($t_{rh}$ = 0.12 Gyr).
These simulations have been already used in Sollima et al. (2015) to test the
bias introduced by the use of multimass King-Michie models in the estimate of
mass and MF.

\begin{figure}
 \includegraphics[width=8.6cm]{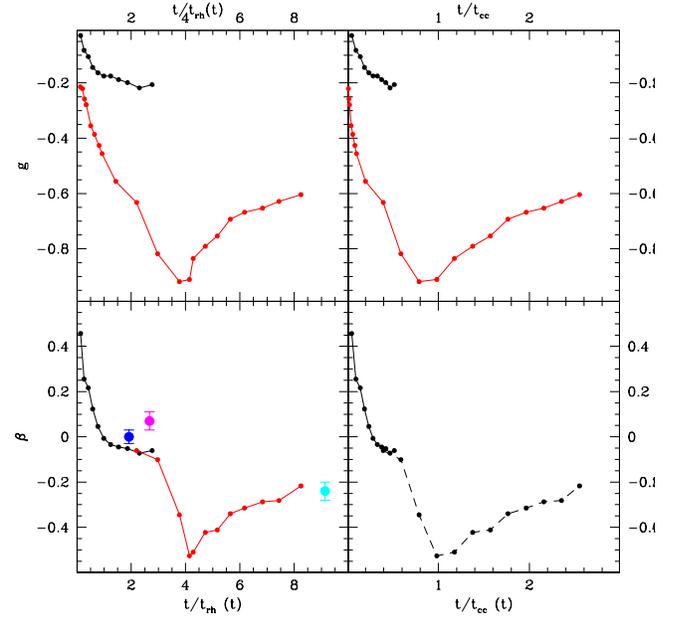}
 \caption{Time evolution of the index $g$ (top panels) and $\beta$ (bottom
 panels) in the two analysed N-body simulations. In left panels time is normalized to the local half-mass relaxation
 time while in right panels it is normalized to the core collapse epoch. The
 locations of the three target GCs in this plane are marked in the bottom-left
 panel as blue (NGC5466), cyan (NGC6218) and magenta (NGC6981) 
 dots (grey dots with increasing darkness are used in the printed version of the
 paper).}
\label{nbody}
\end{figure}

From these simulations we extracted snapshots at different epochs and for each
of them we considered the projected positions on the x-y plane of the unevolved
stars. Stars have been divided in eight evenly spaced mass bins and for each of 
them we calculated the projected radius containing half of their population.
The projected half-mass radii as a function of the star mass have been then fitted
with a power-law whose index $g$ (equivalent to that estimated by Goldsbury et al.
2013) gives an indication of the degree of mass segregation.
The behaviour of this index as a function of time is
shown in the top panels of Fig. \ref{nbody}. In this figure time has been nomalized to two 
characteristic timescale: the half-mass relaxation time $t_{rh}(t)$ (Spitzer
1987) and the core collapse time $t_{cc}$. The index $g$ is $g\sim0$ at the
beginning of both simulation (as expected since simulations stars with no mass
segregation) and then decreases until core collapse. A slow and steady increase
of the $g$ index is apparent in the post-core collapse phase of simulation
W5rh1R8.5. Note that the declining rate of $g$ of
simulation W5rh1R8.5 is steeper than that of simulation W5rh11.5R8.5, 
regardless of the time normalization. This is due to the different efficiency of
two-body relaxation in producing mass segregation in the two simulations.
Indeed, after the same number of half-mass relaxation times, simulation 
W5rh11.5R8.5 lost a significantly larger fraction of its stars with respect to
simulation W5rh1R8.5, because of the strong interaction with the tidal field. 
Thus, it has a flatter MF while mantaining a less concentrated profile. Under
these conditions, the average mass contrast in long-range interactions decreases
reducing the efficency of two-body relaxation. For the same reason, mass
segregation and the radial flows of specific heat leading to core collapse
proceed on different timescales. Summarizing, the behaviour of the mass
segregation index $g$, although linked to the stage of dynamical evolution of the
cluster is highly sensitive to initial conditions like initial MF, strength of
the tidal field, concentration, etc.

For comparison, we considered the fit to the analysed snapshots made in
Sollima et al. (2015) with multimass King-Michie models and calculated the index
$\beta$ using the technique described in Sect. \ref{res_sec}.
The values of $\beta$ are plotted in the bottom panel of Fig. \ref{nbody} as a
function of time. It is interesting to note that the values of $\beta$
derived for the two simulations coincide in the overlap range of time. This
occurs both considering the half-mass relaxation and the core collapse timescale
for normalization. In general, four different stages can be defined in this
plot:
\begin{itemize}
\item{An initial phase (when $t<t_{rh}(t)$) where two-body relaxation is still
not effective in redistributing kinetic energies among stars with different 
masses. In this phase, the actual degree of mass segregation is smaller than the
prediction of multimass King-Michie models still resambling the initial condition 
(decreasing $g$; $\beta>0$);}
\item{An intermediete phase ($t_{rh}(t)<t<t_{cc}$) in which the system sets in an
equilibrium state where its structural variation (size, mass, concentration and
MF) is coupled to a progressive increase of the degree of mass segregation
nicely following the prescription for kinetic energy balance predicted by 
King-Michie models (decreasing $g$; $\beta\sim0$);}
\item{The core-collapse phase where a cusp in the central potential develops.
The presence of this cusp violates the boundary condition of King-Michie models 
at the center ($d\psi/dr(0)\neq0$) making their prediction for mass segregation
underestimated ($\beta<0$);}
\item{The post-core collapse phase where binaries release kinetic energy
during collisions with (mainly massive) stars in the cluster core. This produce
the bounce of the core and a decrease of the degree of mass segregation
(increasing $g$ and $\beta$).}
\end{itemize}

Note that, while the duration of the initial phase is set by the efficiency of
two-body relaxation (thus scaling with the half-mass
relaxation time), the duration of the intermediate phase depends on the many 
parameters affecting the core collapse epoch. In particular, the presence of a 
significant population of hard binaries can delay core collapse by a significant
factor (Gao et al. 1991). So, although such a phase lasts at $t\sim 4~t_{rh}$ in
the considered simulations (run without primordial binaries), its duration could
be several Gyr long in real GCs, representing the most common among their stages of 
dynamical evolution.
The location in this diagram of the three analysed GCs is overplotted to Fig
\ref{nbody}. Note that while NGC5466 and NGC6981 lie in the
intermediate phase region, the negative value of $\beta$ measured in NGC6218,
together with its large value of $t/t_{rh}(t)$ put
this cluster in the post-core collapse phase.

\section{Summary}
\label{sum_sec}

In this paper we showed that the radial distribution of
stars in three Galactic GCs is broadly consistent with the prediction of
analytic King-Michie multimass models. In particular, in two of them (NGC5466 
and NGC6981) we found no appreciable difference in the dependence of 
the characteristic radii on stellar mass.
This result confirms what already found in Sollima et al. (2015),
who however based their analysis on a set of N-body
simulations and focussed only on the accuracy in the mass and MF estimate.

In the commonly used formulation by Gunn \& Griffin (1979), these models 
assume that the exponent in the lowered-Maxwellian distribution
function (see eq. \ref{eq_df}) are directly proportional to the stellar
mass. Note that, because of the tidal truncation in energy, this condition
differs from kinetic energy equipartition. 
Indeed, the actual squared velocity dispersions of the 
different mass groups in this model do not follow a 
$\sigma\propto m^{-0.5}$ relation neither globally nor
at any radius.
On the other hand, the resulting behaviour of the $\sigma-m$ relation is closer 
to equipartition in the central region than in the outer parts.
Note that the assumption made by Gunn \& Griffin (1979) is arbitrary and other choices
are equally justified (see e.g. Merritt 1981).
In spite of this, the results presented here show that these models provide a
good representation of real GCs, although the choice of the (uncertain) 
recipies to account for the
fraction of retained remnants and the characteristics of the binary
population have significant effects.
The comparison with N-body simulation
spanning the entire evolution of a simulated star cluster under two different
initial conditions indicates that this agreement is expected during a time
interval between the half-mass relaxation time and the core collapse epoch,
regardless of the initial conditions. Such a time interval is expected to
constitute a significant portion of the evolution of GCs being probably the most
common stage experienced by present-day GCs. 
On the other hand, significant differences between
King-Michie models and GCs are instead
expected when the cluster is still dynamically young ($t<t_{rh}$) or near and after 
core collapse. In this regard, the parameter $\beta$, defined as the difference
between the observed and predicted power-law indices of the characteristic
radius-stellar mass relation, has been found to be useful in distinguishing the
dynamical stage of an observed cluster, being almost insensitive to initial
conditions. 

An notable case is represented by the GC NGC6218
whose massive stars appears to be significantly more segregated than what 
predicted by models. This is indicated by the negative value of the 
parameter $\beta$ measured in
this cluster ($\beta=-0.24\pm0.04$) at odds with those estimated in the other
three GCs of our sample ($\beta\sim 0$).
Note that this cluster has a flat MF ($\alpha=-0.36$; suggesting a strong mass 
loss history; de Marchi, Pulone \& Paresce 2006; Sollima et al. 2012) and a 
short present-day
half-mass relaxation time ($t_{rh}(t)\sim1.4~Gyr$). It is therefore possible 
that it could have already experienced core collapse in
the past.

\section*{Acknowledgments}

We warmly thank the anonymous referee for his/her helpful comments and
suggestions.
AS acknowledges the PRIN INAF 2011 "Multiple populations in globular
clusters: their role in the Galaxy assembly" (PI E. Carretta) and the PRIN INAF
2014 ``Probing the internal dynamics of globular clusters. The first,
comprehensive radial mapping of individual star kinematics with the
generation of multi-object spectrographs" (PI L. Origlia). 
ED and CP acknowledge support from the European 
Research Council (ERC-2010-AdG-267675, COSMIC-LAB).

\label{lastpage}

\end{document}